\documentclass[aps,prl,twocolumn,showpacs,preprintnumbers,amsmath,amssymb]{revtex4-1}
\usepackage{latexsym}
\usepackage{graphicx}
\usepackage{graphics}
\usepackage[T1]{fontenc}
\usepackage{multirow}
\usepackage{changes}

\newcommand{\ketbra}[2]{\mbox{$ |#1\rangle \langle #2 |$}}

\newcommand{\ket}[1]{\mbox{$ | #1 \rangle $}}

\begin{document}

\title{Narrow inhomogeneous distribution of spin-active emitters in silicon carbide}

\author{Roland Nagy$^{1,2}$, Durga Bhaktavatsala Rao Dasari$^2$, Charles Babin$^2$, Di Liu$^2$, Vadim Vorobyov$^2$, Matthias Niethammer$^2$, Matthias Widmann$^2$, Tobias Linkewitz$^2$, Izel Gediz$^2$, Rainer St\"ohr$^2$, Heiko B. Weber$^3$, Takeshi Ohshima$^4$, Misagh Ghezellou$^5$, Nguyen Tien Son$^5$, Jawad Ul-Hassan$^5$}
\author{Florian Kaiser$^2$}
\email{f.kaiser@pi3.uni-stuttgart.de}
\author{J\"org Wrachtrup$^2$}

\affiliation{$^1$Department Elektrotechnik-Elektronik-Informationstechnik (EEI), Friedrich-Alexander-Universit\"at Erlangen-N\"urnberg (FAU), 91058 Erlangen (Germany)\\
$^2$3rd Institute of Physics, IQST, and Research Center SCoPE, University of Stuttgart, 70569 Stuttgart (Germany)\\
$^3$Department of Physics, Friedrich-Alexander-Universit\"at Erlangen-N\"urnberg (FAU), 91058 Erlangen (Germany)\\
$^4$National Institutes for Quantum and Radiological Science and Technology, Takasaki, Gunma 370- 1292 (Japan)\\
$^5$Department of Physics, Chemistry and Biology, Link\"oping University, SE-58183 Link\"oping (Sweden)}

\begin{abstract}
Optically active solid-state spin registers have demonstrated their unique potential in quantum computing, communication and sensing. Realizing scalability and increasing application complexity requires entangling multiple individual systems, e.g. via photon interference in an optical network. However, most solid-state emitters show relatively broad spectral distributions, which hinders optical interference experiments.
Here, we demonstrate that silicon vacancy centres in semiconductor silicon carbide (SiC) provide a remarkably small natural distribution of their optical absorption/emission lines despite an elevated defect concentration of $\approx 0.43\,\rm \mu m^{-3}$. In particular, without any external tuning mechanism, we show that only 13 defects have to be investigated until at least two optical lines overlap within the lifetime-limited linewidth.
Moreover, we identify emitters with overlapping emission profiles within diffraction limited excitation spots, for which we introduce simplified schemes for generation of computationally-relevant Greenberger-Horne-Zeilinger (GHZ) and cluster states.
Our results underline the potential of the CMOS-compatible SiC platform toward realizing networked quantum technology applications.
\end{abstract}

\maketitle
Well-controlled interference at the single and multi-photon levels is essential for quantum information~\cite{KLM_KLM_Nature_2001}. Original studies on the fundamental behaviour of quantum light have now turned into applications in the fields of quantum optical sensing~\cite{Lemos.2014, Pirandola.2018,Kaiser_Sesnsing_LSA_2018}, communication~\cite{Liao.2017, Simon.2017, Chen_QKD_PRL_2020} and computation~\cite{OBrien.2007, Zhong_76Photons_Science_2020}. Unique possibilities arise with light-matter interfaces in which quantum systems coherently couple photons and internal degrees of freedom, e.g. a spin-based quantum memory~\cite{DLCZ_DLCZ_Nature_2001,Yuan_Repeater_Nature_2008,Dolde_Coupling_NatPhys_2013,Blatt_Entanglement_PRL_2013,Yang.2016, Humphreys_Deterministic_Nature_2018, Bock_Ions_NatComm_2018,  Hacker_Cats_NatPhoton_2019}. With such an interface, strong (quasi-nonlinear) interactions between multiple photons can be mediated, which is crucial for boosting efficiencies in memory-enhanced quantum communication~\cite{Bhaskar_memory_2020} and measurement-based quantum computing~\cite{Briegel_MBQC_NatPhys_2009}. The full potential of those quantum technologies will be unleashed by realizing large-scale architectures based on establishing entanglement amongst multiple systems via photonic interference in an optical network~\cite{Blatt_Entanglement_PRL_2013,Rosenfeld_Loophole_PRL_2017,Humphreys_Deterministic_Nature_2018}. Potentially, optical loss in such networks can be overcome by encoding information into cluster states~\cite{Buterakos_Deterministic_PRX_2017, Tiurev_Cluster_arXiv_2020}.

Reaching this goal requires highly indistinguishable photons from distinct spin-active emitters, which has remained a major challenge for solid-state systems, in particular due to variations in their local crystal environments~\cite{Batalov.2009, Rogers.2014}.
Landmark multi-photon interference experiments have been achieved by overlapping the emission of two nitrogen-vacancy (NV) color centres in diamond~\cite{Bernien.2012}. Via sophisticated electric control structures, optical properties have been tuned, permitting to mediate entanglement among distant NV centre spins via photonic interference~\cite{Humphreys_Deterministic_Nature_2018}. Further scalability is challenged by the NV centre's notorious sensitivity to local charge fluctuations. Recently, strong interest has been gained by inversion-symmetry color centres in diamond as sources of indistinguishable photons, with one work showing that an investigation of 20 silicon vacancy centres resulted in 11 overlapping emission lines~\cite{Rogers.2014}. Due to the system's elevated strain sensitivity, results were reported only on deep bulk defects in an ultra-low strain diamond substrate, which does not adequately represent the conditions in nanofabricated devices~\cite{Wan_Integration_2020}.

To additionally overcome some of the intrinsic drawbacks of the diamond platform (material availability, fabrication, Fermi level control), we investigate here the silicon mono-vacancy center ($\rm V_{Si}$) in the 4H polytype of silicon carbide (4H-SiC)~\cite{Fuchs_SiC_NatComm_2015, Widmann_coherent_2015, Nagy.2018, Chen.2019}. Structurally, there are two distinct $\rm V_{Si}$ centers, formed by a missing silicon atom on a lattice site with hexagonal (h-$\rm V_{Si}$) or cubic (c-$\rm V_{Si}$) symmetry~\cite{Sorman_SiC_PRB_2000, ErikJanzen.2009}.
Due to identical ground and excited state wavefunction symmetry, both defects show similarly stable optical properties than their diamond counterparts~\cite{Nagy.2019, Udvarhelyi_Stable_PRAppl_2019, Banks_V2_2019, Morioka.2020, Udvarhelyi_Vibronic_2020}, however spin coherences are maintained up to room temperature~\cite{Widmann_coherent_2015}.
This work focuses on the h-$\rm V_{Si}$ center that shows a strong zero-phonon line (ZPL) at 861.6\,nm. Recent research demonstrated that individual h-$\rm V_{Si}$ centers combine promising spectral properties with a high-coherence electron spin~\cite{Nagy.2019}. Additionally, a high-fidelity spin-photon interface was demonstrated that showed that single h-$\rm V_{Si}$ centers can be used for generation of spin-photon entanglement and frequency-encoded photonic cluster states~\cite{Morioka.2020}.

In the following, we provide a detailed investigation of the spectral properties of 50 randomly chosen near-surface h-$\rm V_{Si}$ centers. Our results show a remarkably small natural distribution of the optical absorption/emission lines. In particular, in the investigated sample, we find that merely 13 defects have to be investigated until at least one pair of overlapping emission lines is found with greater than 50\% probability.
This underlines the system's potential toward scaling up quantum information platforms via photonic interference~\cite{Calusine.2014, Englund.2010}.

\begin{figure}
\centering
\includegraphics[width=\columnwidth]{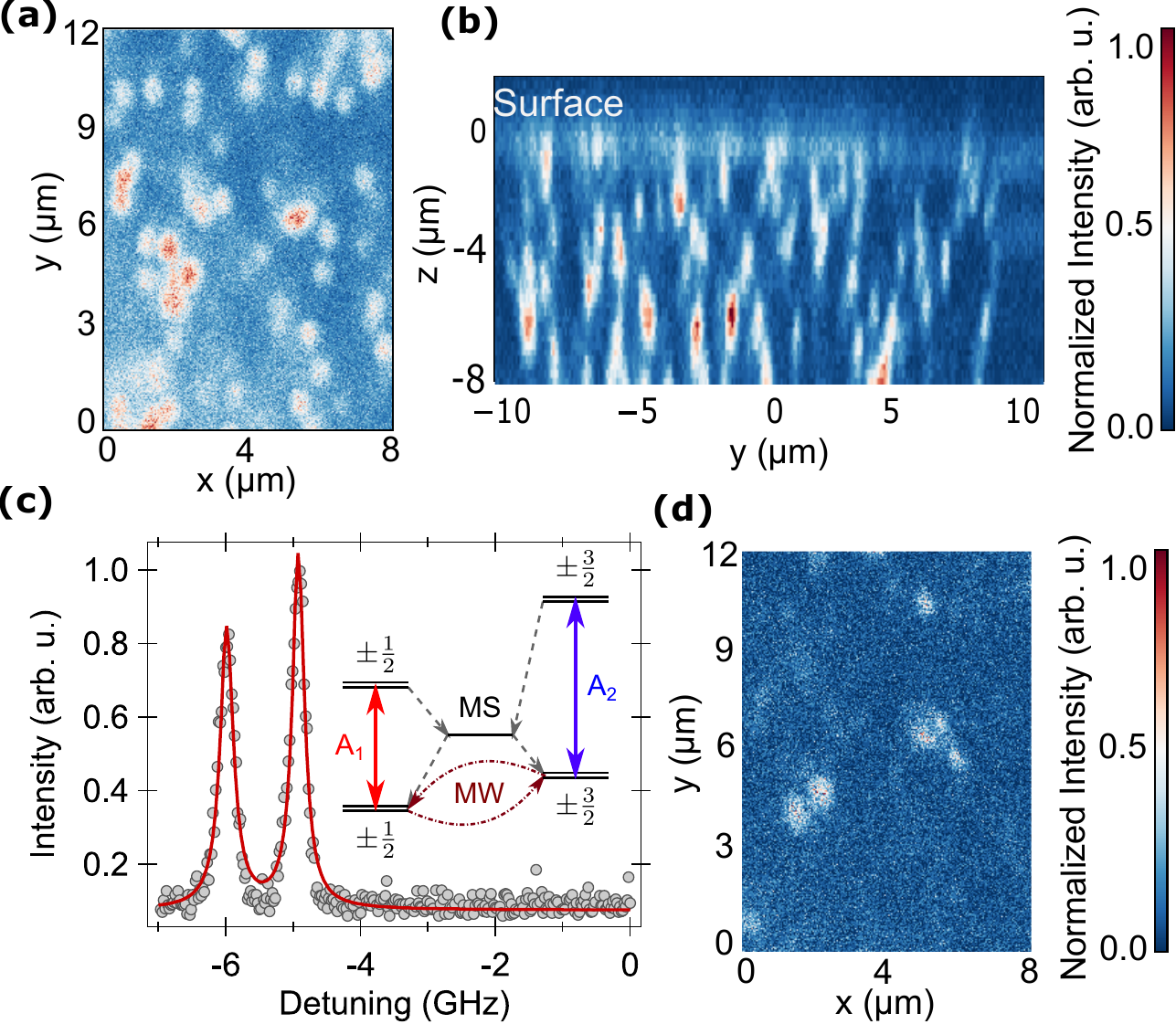}
\caption{Optical properties of h-$\rm V_{Si}$ centres. (a) Confocal scan in the xy-plane using above-resonant excitation at 730\,nm. Bright spots indicate h-$\rm V_{Si}$ centres. (b) Depth scan (yz-coordinates) using above-resonant excitation at 730\,nm. Defects are found throughout the entire sample. In this work, we focus exclusively on defects within 1 micrometer surface proximity.
(c) Representative resonant excitation photoluminescence excitation scan on a single h-$\rm V_{Si}$ centre. The two characteristic absorption lines are associated with the optical transitions in the spin-$\tfrac{1}{2}$ and spin-$\tfrac{3}{2}$ subspaces (see inset). The fit to the data is based on a double-Lorentzian function. (d) Confocal scan in the xy-plane using resonant excitation on the A$_2$ optical transition of the defect identified in (c). Multiple bright spots appear, corroborating identical absorption energies of various emitters.}
\label{fig:Confocal}
\end{figure}

In this work, we take advantage of previously-developed experimental methods that are described in detail in Ref.~\cite{Nagy.2019}. As opposed to this work, we investigate here another 4H-SiC crystal with a $\approx 10\times$ higher defect center concentration, thus permitting detailed statistical analysis.
Individual h-$\rm V_{Si}$ centers in an isotopically ultrapure slightly n-type 4H-SiC sample are addressed via a home-built confocal microscope that operates at a temperature of $T=5\,\rm K$ (see Ref.~\cite{Nagy.2019}). To identify single defects, we perform confocal scans using off-resonant excitation at 730\,nm, while collecting red-shifted phonon sideband (PSB) fluorescence ($880 - 1000\,\rm nm$), see \figurename~\ref{fig:Confocal}(a, b). By fitting an isolated emitter in \figurename~\ref{fig:Confocal}(b) with a two-dimensional Gaussian function, we deduce that our optical system has an axial resolution of $\approx 1.22\,\rm \mu m$. Subsequently counting the defects in \figurename~\ref{fig:Confocal}(a), we then estimate an average color center density of $\approx 0.43\,\rm \mu m^{-3}$. For the remainder of this paper, we focus exclusively on near-surface color centres ($< 1\,\rm \mu m$ depth) to (at least partially) mimic some conditions in nanofabricated environments. However, we mention that integration into actual nanophotonic structures sets additional challenges regarding surface charge fluctuations and strain. In a second step, we identify the optical absorption/emission lines of individual h-$\rm V_{Si}$ centers. To this end, we use resonant excitation using a wavelength-tunable 861.6\,nm diode laser (Toptica DLC DL pro), while collecting PSB emission. To avoid dark states due to optical pumping, we continuously mix the spin ground states by additionally providing microwave (MW) excitation at 5\,MHz through a nearby copper wire~\cite{Nagy.2019}. \figurename~\ref{fig:Confocal}(c) presents exemplary photoluminescence excitation data for one h-$\rm V_{Si}$ center, showing the well-known absorption/emission lines that are separated by $\approx \rm 1\,GHz$. The low-frequency line ($\rm A_1$) is associated with the spin-conserving transition linking the spin-$\tfrac{1}{2}$ subspaces of the ground and excited states, while the high-frequency transition ($\rm A_2$) belongs to the spin-$\tfrac{3}{2}$ subspace (\figurename~\ref{fig:Confocal}(c) inset). To highlight the remarkably small spectral distribution between the emission lines of multiple h-$\rm V_{Si}$ centers, we perform a resonant-excitation confocal scan in the same area as in \figurename~\ref{fig:Confocal}(a). For the scan, we fix the resonant excitation laser at the $\rm A_2$ line of the previously identified  h-$\rm V_{Si}$ center. The image in
\figurename~\ref{fig:Confocal}(d) shows multiple bright spots, which corroborates that a large fraction of h-$\rm V_{Si}$ centers shows matching spectral lines (either $\rm A_1$ or $\rm A_2$).

\begin{figure}
\centering
\includegraphics[width=\columnwidth]{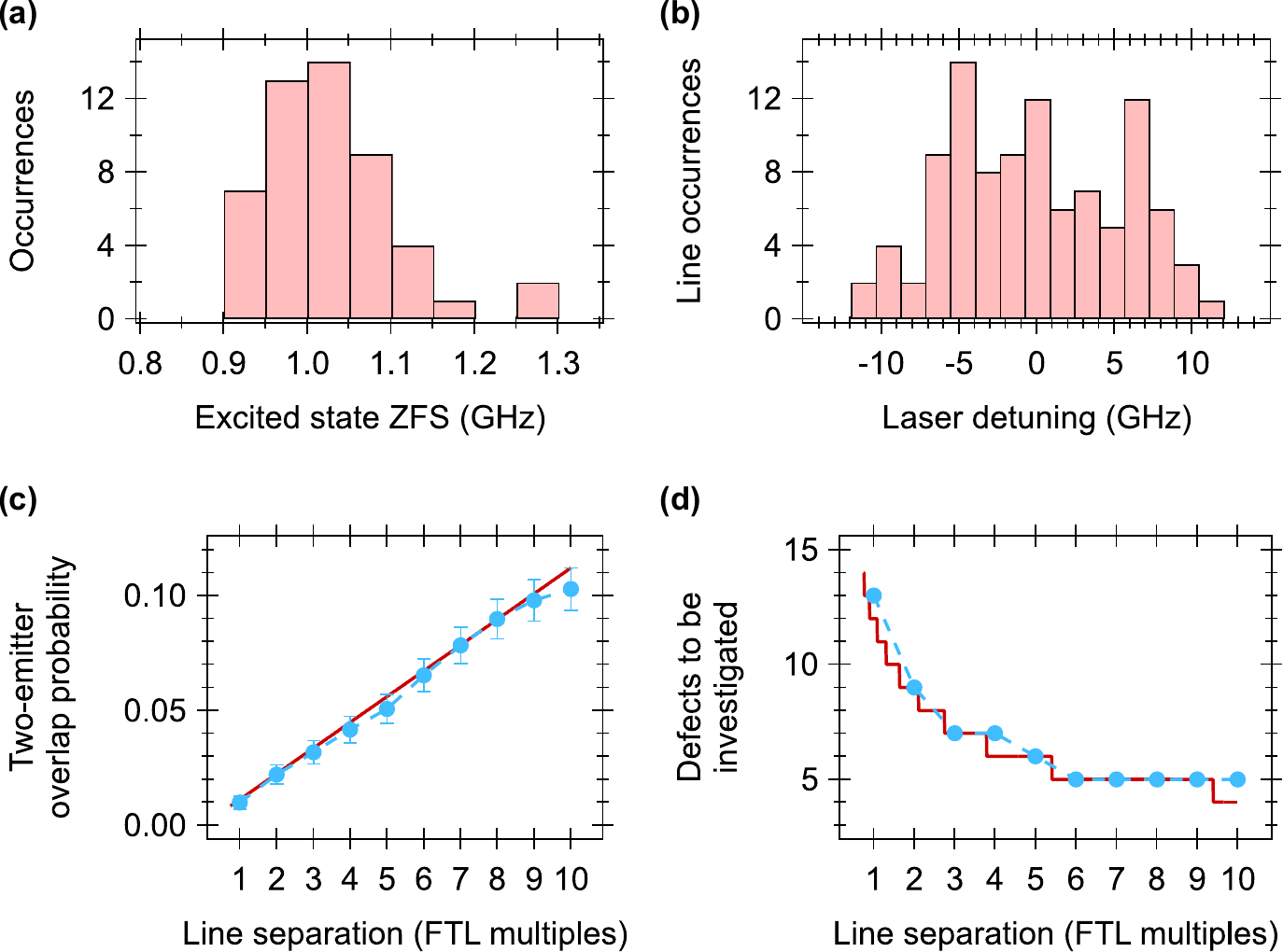}
\caption{Statistics on the spectral distribution of 50 near-surface h-$\rm V_{Si}$ centres in 4H-SiC. (a) Distribution of the measured excited stated ZFSs.
 (b) Histogram of the A$_{1}$ and A$_{2}$ absorption frequencies. The frequency detuning at 0 GHz corresponds to an absolute frequency of 347.94059\,THz.
(c) Probability that two randomly chosen h-$\rm V_{Si}$ centres show at least one pair of overlapping lines in a given frequency range. The linear fit to the data considers a homogeneous distribution of absorption lines. For small line separations, the experimental data are above the fit due to the bunching behaviour (see \figurename~\ref{fig:Statistics}(a)). Error bars represent one standard deviation. (d) Number of defects to be investigated until at least one pair of overlapping emission lines is observed with probability $>50\%$. As in (c), the fit to the data assumes a homogeneous line distribution, which explains the deviation for smaller line separations.}
\label{fig:Statistics}
\end{figure}

To gain further insights into the spectral distribution, we investigate the absorption lines of 50 h-$\rm V_{Si}$ centers. On all defects, we perform resonant photoluminescence excitation studies (identical to the procedure in \figurename~\ref{fig:Confocal}(c)) and reference the laser frequency with a wavemeter (HighFinesse WS8-30; 30\,MHz absolute accuracy, 1\,MHz resolution). \figurename~\ref{fig:Statistics}(a) shows the frequency separation between the $\rm A_1$ and $\rm A_2$ transitions. Due to the h-$\rm V_{Si}$ centre's small ground state splitting of about 5\,MHz, the line separation is mainly determined by the excited state zero field splitting (ZFS). The histogram shows a central value of 1.027\,GHz with a remarkably small distribution of 75\,MHz (one standard deviation).
 \figurename~\ref{fig:Statistics}(b) shows the distribution of the central frequencies of the $\rm A_1$ and $\rm A_2$ lines. Quite remarkably, we find that all absorption lines are found within a narrow spectral range, covering only $\pm 10\rm\,GHz$, which is comparable to naturally occurring deep-bulk silicon vacancy centers in diamond~\cite{Rogers.2014}. As shown in the Supplementary Information, no significant variation of the distribution is found when analysing defects in separated regions, thus corroborating homogeneous properties throughout the sample.
We also compare our results to previous results obtained in 4H-SiC with $\approx 10 \times$ lower h-$\rm V_{Si}$ center densities~\cite{Nagy.2019,Morioka.2020}. Here, inhomogeneous distributions of $2-3\,\rm GHz$ were measured, however the studied sample size was relatively small due to the limited number of available defects. In future work, it may be interesting to study the relationship between defect density and inhomogeneous distribution, which could have direct impact on the optimal defect density for ensemble-based quantum applications.

High-quality interference of photons from two distinct emitters requires a good spectral overlap, i.e. the frequency separation between the emission lines must be within the defects' lifetime-limited emission linewidths. Considering that the excited-state lifetime of the h-$\rm V_{Si}$ centre is $\tau \approx 5.5\rm\,ns$~\cite{Nagy.2018} results in a Fourier transform limited emission linewidth of $\Gamma = \left( 2 \pi \, \tau \right)^{-1} \approx 29\rm\,MHz$. Similarly to Ref.~\cite{Rogers.2014}, we find that out of the 50 investigated defects, 12 pairs show emission lines that are separated by less than one lifetime-limited emission linewidth. However, we note that such an analysis is prone to misinterpretation, as the number of overlapping emitters increases approximately quadratically with the number of investigated pairs. For this reason, we perform additional statistical analysis. In \figurename~\ref{fig:Statistics}(c) we show the probability that two randomly chosen emitters have overlapping emission lines in a given frequency range. Experimentally, we infer a probability of $1.0 \pm 0.3\%$ for an overlap within one lifetime-limited linewidth $(29\rm\,MHz)$. Additionally, the linear fit to the data shows a slope of 1.12(3)\% per lifetime-limited linewidth. This is comparable to deep-bulk silicon vacancies in diamond (slope of 5.8\% per lifetime-limited linewidth, deduced from Ref.~\cite{Rogers.2014}), and slopes of about 10\% per lifetime-limited linewidth can be expected for the previously studied 4H-SiC samples with low densities of h-$\rm V_{Si}$ centres~\cite{Nagy.2019,Morioka.2020}. In addition, we mention that unlike diamond defects, all h-$\rm V_{Si}$ centres occur with the same optical dipole orientation~\cite{Nagy.2018, Nagy.2019}, thus facilitating the necessary polarization mode overlap in interference experiments, and improving the yield for coupling colour centres to nanophotonic resonators.

In view of system scalability, arguably the most critical parameter is how many defect centres have to be investigated until at least one pair of overlapping lines is identified.
In direct analogy to the popular birthday paradox~\cite{Frank_Bday_AMS_1964} it turns out that the number is remarkably small. As shown in \figurename~\ref{fig:Statistics}(d), we find that only 13 h-$\rm V_{Si}$ centres have to be investigated to have more than 50\% chance to find at least one pair of overlapping emission lines.
We note that the number of h-$\rm V_{Si}$ centres to be investigated is expected to reduce drastically with their integration into cavities, due to the lifetime shortening~\cite{Calusine.2014, Englund.2010, Bracher.2017,  Magyar.2014, Lukin.2020b} and the associated increase in the lifetime-limited linewidth. Further, deterministic overlap can be achieved using additional external tuning mechanisms, e.g. based on electrical fields and/or strain~\cite{lasCasas.2017, Falk.2014, Ruhl.2020, Lukin_spectral_2020}. These effects can also compensate for the slightly increased linewidths observed in this sample, i.e., the average full width at half maximum is found to be 316\,MHz with a standard deviation of 122\,MHz. We attribute this to global properties of the 4H-SiC sample (e.g., an elevated Fermi level) rather than the increased defect density. This statement is corroborated by the fact that we do not observe a change in the linewidth distribution as a function of the local defect density (see Supplementary Information).

\begin{figure}
\centering
\includegraphics[width=\columnwidth]{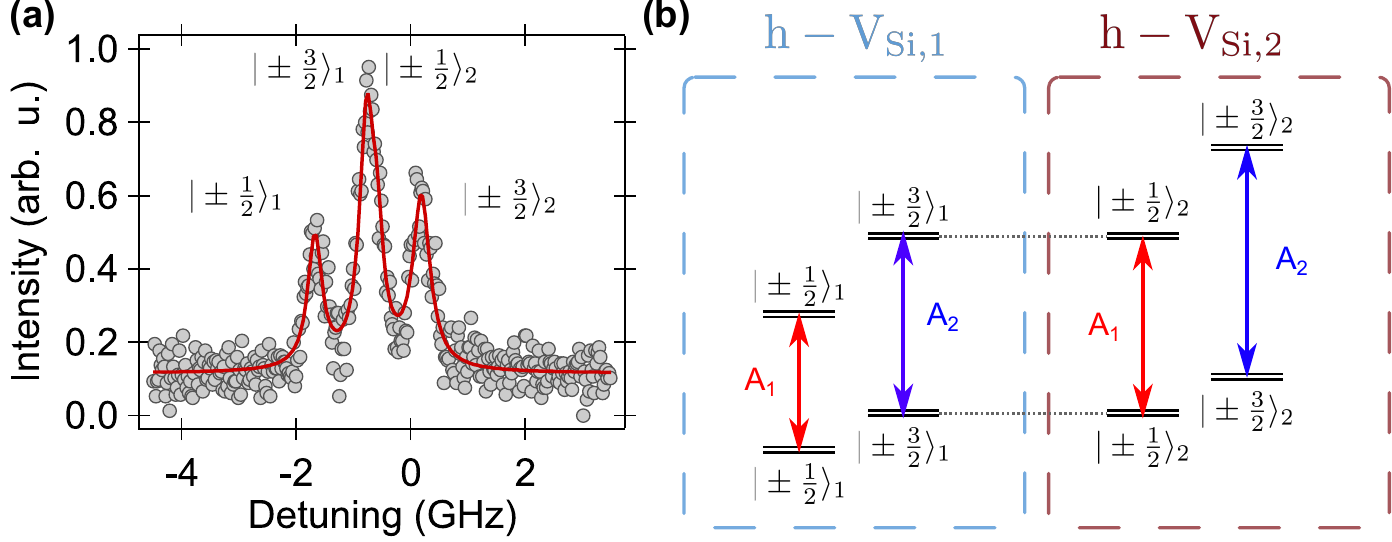}
\caption{Overlapping optical emission of a h-$\rm V_{Si}$ pair located within one diffraction limited spot. (a) Resonant photoluminescence excitation measurement on a h-$\rm V_{Si}$ pair. The central peak in the absorption spectrum is assigned to A$_2$ and A$_1$ optical lines of the first and second defect, respectively. (b) The observed absorption lines are assigned to different spin-dependent optical transitions in the spin-$\tfrac{1}{2}$ and spin-$\tfrac{3}{2}$ subspaces. Such a system could be used for photon-mediated spin-spin entanglement generation.}
\label{fig:SpecialCase}
\end{figure}

We mention further that due to the relatively high defect density in our 4H-SiC sample, we actually regularly identify multiple h-$\rm V_{Si}$ centres within a single diffraction limited confocal excitation spot. A particularly interesting resonant photoluminiscence excitation spectrum of a h-$\rm V_{Si}$ pair is shown in \figurename~\ref{fig:SpecialCase}(a), presenting three optical transitions. As depicted in \figurename~\ref{fig:SpecialCase}(b), we assign the lowest-frequency peak to the A$_1$ optical transition in the spin-$\tfrac{1}{2}$ subspace of the first defect centre (h-$\rm V_{Si,1}$), while the highest-frequency peak is attributed to the A$_2$ transition in the spin-$\tfrac{3}{2}$ subspace of the second defect (h-$\rm V_{Si,2}$). The central peak comprises the overlapping A$_2$ and A$_1$ optical lines of the first and second defect, respectively. Although it is unlikely that both defect centres show dipolar spin-spin coupling~\cite{Dolde_Coupling_NatPhys_2013}, the spectral overlap of two non-identical transitions can facilitate generation of spin-spin entanglement and spin-optical cluster states. In particular, as both emitters are located within the diffraction limit, all optical paths overlap and are therefore automatically phase-stable. This relaxes considerably the constraints for complex stabilisation systems, which is currently a major source of decoherence~\cite{Humphreys_Deterministic_Nature_2018}. An important feature of such a two-defect system arises when using an optical excitation pulse that is resonant with the central absorption line. Similarly to previous work~\cite{Humphreys_Deterministic_Nature_2018}, this leads to three distinguishable outcomes: (i) zero-photon emission if the collective ground-state spin is in the state $\ket{\pm 1/2}_1 \ket{ \pm 3/2}_2$, (ii) single-photon emission for $\ket{\pm 1/2}_1 \ket{ \pm 1/2}_2$ or $\ket{\pm 3/2}_1 \ket{ \pm 3/2}_2$ and (iii) two-photon emission if the ground-state spin is in the state $\ket{\pm 3/2}_1 \ket{ \pm 1/2}_2$. In all spin states, the subscript denotes the defect index. The most interesting case is (ii), in which the emission of a single photon projects both defects into identical spin states. Provided that the optical emission is resonant, i.e., in the zero-photon line, it is further possible to generate coherent superposition states among pairs of spins. Using this spin-optical feature, we will now introduce the generation of GHZ states conditioned on photon-number resolved measurements.

\begin{figure}
\centering
\includegraphics[width=\columnwidth]{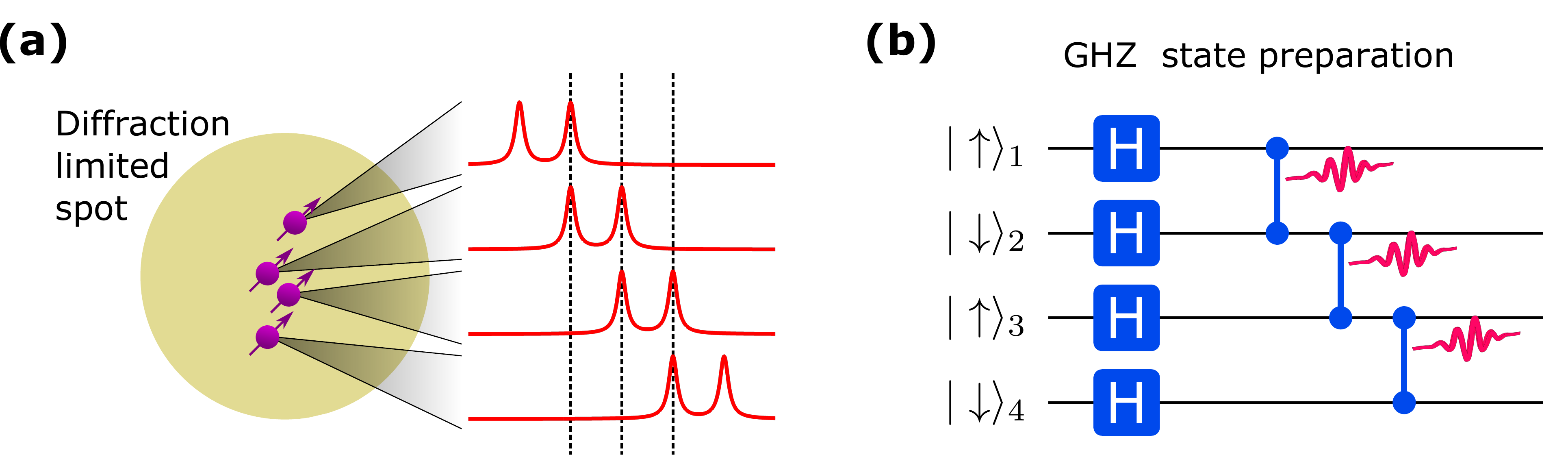}
\caption{Generation of network-relevant entangled spin-states states. (a) Considered multi-emitter system, comprising four h-$\rm V_{Si}$ centres within a diffraction limited spot. All defects show pairwise spectral overlap. (b) Protocol for generation of GHZ states. The vertical lines and the photon (undulated lines) describe the conditional emission of a photon only when the spin-states of the connected defects are identical.}
\label{fig:GHZcluster}
\end{figure}

To simplify the reading, we represent the ground state of the h-$\rm V_{Si}$ centre as a two-level qubit, i.e., the spin-$\tfrac{1}{2}$ subspace is summarised as $\ket{\uparrow}$ and the spin-$\tfrac{3}{2}$ subspace as $\ket{\downarrow}$. As shown in \figurename~\ref{fig:GHZcluster}(a), we consider an array of four emitters that show a pairwise overlap of the A$_1$ and A$_2$ transitions. Although identifying this exact spectral arrangement in a single spot may be somewhat unlikely, several separated defect clusters may be tuned into resonances using Stark shift control~\cite{Ruhl.2020, Lukin_spectral_2020}.
To initialise the spin ground states of the four defects efficiently, we can use laser excitation that is resonant with the A$_2$ transition of defect 1 and, at the same time, resonant with the A$_1$ transition of defect 2. Optical pumping eventually results in a dark state~\cite{Nagy.2019}, $\ket{\uparrow}_1 \ket{\downarrow}_2$. Similarly, defects 3 and 4 can be initialised, resulting in the state $\ket{\psi_{0, {\rm GHZ}}} = \ket{\uparrow}_1 \ket{\downarrow}_2 \ket{\uparrow}_3 \ket{\downarrow}_4$. The subsequent protocol is shown in \figurename~\ref{fig:GHZcluster}(b). We propose to create an equal superposition of all spin states using a common MW $\pi/2$-pulse (Hadamard gate), resulting in: 
$\ket{\psi_{{\rm H,\,GHZ}}} =\ket{+}_1 \ket{-}_2 \ket{+}_3 \ket{-}_4$, with $\ket{\pm}_i = \left( \ket{\uparrow}_i \pm \ket{\downarrow}_i \right)/\sqrt{2}$ ($i=1,\,2,\,3,\,4$). Using the optical feature described above, resonant excitation in the spectral overlap region between defects 1 and 2, and conditioning on the emission of a single photon ($\nu$), projects the above state to:
\begin{equation}
    \ket{\psi_{\nu, {\rm GHZ}}} = \tfrac{1}{\sqrt{2}} \left( \ket{\uparrow}_1 \ket{\uparrow}_2 - \ket{\downarrow}_1 \ket{\downarrow}_2  \right) \ket{+}_3 \ket{-}_4.
\end{equation}
Repeating the same optical excitation and conditioning on single photon emission subsequently for defects 2 and 3, and thereafter for defects 3 and 4 projects the final state into:
\begin{equation}
    \ket{\psi_{{\rm GHZ}}} = \frac{ \ket{\uparrow}_1 \ket{\uparrow}_2 \ket{\uparrow}_3 \ket{\uparrow}_4 + \ket{\downarrow}_1 \ket{\downarrow}_2 \ket{\downarrow}_3 \ket{\downarrow}_4}{\sqrt{2}}.
\end{equation}
Note that the production of this state relies on single-photon detection events. Thus, any deviation from unity photon detection efficiency $\eta$ reduces the state fidelity as single-photon loss on a two-photon emission event leads to false positives. E.g., in the case of two emitters, detecting a single photon leads to a mixed state: $\rho = p \ketbra{\Phi^+_{1,2}}{\Phi^+_{1,2}} + (1-p)\ketbra{\downarrow_1\uparrow_2}{\downarrow_1\uparrow_2}$, where $p =1/(3-2\eta)$ determines both the fidelity of the two-emitter spin state, and $\ket{\Phi^+_{1,2}}$ is the two-spin Bell state. Considering state-of-the-art detection efficiencies of $85\%$~\cite{Bhaskar_memory_2020}, we find GHZ state generation fidelities of $F_2 \sim 0.8,\,F_3 \sim 0.6,\,F_4 \sim 0.5$, for $n=2,\,3,\,4$ spins, respectively. In the Supplementary Information, we show additionally, how the spin-$\tfrac{3}{2}$ nature of $\rm V_{Si}$ centres can be used advantageously for the generation of multi-spin cluster states.

We have shown a large-scale statistical analysis of the spectral properties of h-$\rm V_{Si}$ centres in 4H-SiC. Even in the absence of external tuning mechanisms, we inferred a remarkably high spectral overlap between multiple single emitters. Critically, solely 13 defect centres have to be investigated until one pair of overlapping emission lines is found within the lifetime-limited emission linewidth.
Developing a scalable networked quantum technology platform is further facilitated by the fact that all h-$\rm V_{Si}$ centres show identical spin and optical dipole orientations~\cite{Nagy.2018, Nagy.2019}, long spin coherence times are observed up to room temperature~\cite{Widmann_coherent_2015}, and that the optical emission is nearly transform limited~\cite{Morioka.2020}. 
We outlined how multiple h-$\rm V_{Si}$ centres can be orchestrated to generate GHZ, which are relevant for networked quantum technologies.
In this regard, the SiC platform offers a plethora of additional benefits~\cite{Son_Developping_2020}, such as the integration of defects into p-i-n structures for charge state control~\cite{Anderson_pin_2019}, nonlinear on-chip frequency conversion~\cite{Lukin.2020b}, the demonstrated ability to produce high-quality resonators~\cite{Guidry_OPO_2020}, and the potential for large-scale nuclear spin registers based on the diatomic crystal nature~\cite{Nagy.2019, Bourassa_nuclear_2020}.
Therefore, optically active spin in SiC, especially the h-$\rm V_{Si}$ centre, have a great potential to enable next-generation networked quantum technologies.

\section*{Supplementary Material}

The Supplementary Material comprises derivations of the equations given in the main text. Further, we provide additional data analysis on the statistics of the optical linewidths, as well as investigations in areas with different defect densities.
We present a proposal toward the generation of spin-cluster states. Finally, we outline the mathematics toward inferring the fidelity of generated the multi-spin states in the presence of photon loss. Moreover, we sketch a scheme for generation of spin-controlled multi-photon states.

$\,$\\
$\,$\\
R.N. acknowledges support by the Carl-Zeiss-Stiftung.
F.K. and J.W. acknowledge the EU-FET Flagship on Quantum Technologies through the projects ASTERIQS (grant agreement ID: 820394) and QIA (grant agreement ID: 820445), as well as the German Federal Ministry of Education and Research (BMBF) for the project Q.Link.X (grant agreement 16KIS0867).
J.W. acknowledges support by the European Research Council (ERC) grant SMel, the European Commission Marie Curie ETN ``QuSCo'' (grant agreement No 765267), the Max Planck Society, the Humboldt Foundation, and the German Research Foundation (SPP 1601).
T.O. acknowledges the JSPS KAKENHI (grant Nos. 18H03770 and 20H00355).
N.T.S. acknowledges the Swedish Research Council (grant No. VR 2016-04068). J.U.H. acknowledges the Swedish Energy Agency (grant No.  43611-1) and Swedish Research Council (grant No. 2020-05444). N.T.S. and J.U.H. thank the EU H2020 project QuanTELCO (Grant No. 862721) and the Knut and Alice Wallenberg Foundation (grant No. KAW 2018.0071).

\section*{Data availability}
The data that support the findings of this study are available from the corresponding author upon reasonable request.


\end{document}